\documentclass[twocolumn,showpacs,preprintnumbers,amsmath,amssymb,prl]{revtex4}

\usepackage{graphicx}
\usepackage{dcolumn}
\usepackage{bm}
\usepackage{amsmath}

\usepackage{braket}

\def\degC{\kern-.2em\r{}\kern-.3em C}

\begin{document}

\preprint{******}

\title{Quantum memory of a squeezed vacuum for arbitrary frequency sidebands}

\author{Manabu Arikawa$^{1}$, Kazuhito Honda$^{1}$, Daisuke Akamatsu$^{1}$, \\ Satoshi Nagatsuka$^{1}$, Akira Furusawa$^{2}$, and Mikio Kozuma$^{1,3}$}

\affiliation{%
$^{1}$Department of Physics, Tokyo Institute of Technology, 2-12-1 Okayama, Meguro-ku, Tokyo 152-8550, Japan}

\affiliation{%
$^{2}$Department of Applied Physics, School of Engineering, The University of Tokyo, 7-3-1 Hongo, Bunkyo-ku, Tokyo 113-8656, Japan}
\affiliation{%
$^{3}$CREST, Japan Science and Technology Agency, 1-9-9 Yaesu, Chuo-ku, Tokyo 103-0028, Japan}%

\date{\today}

\begin{abstract}
We have developed a quantum memory technique that is completely compatible with current quantum information processing for continuous variables of light, where arbitrary frequency sidebands of a squeezed vacuum can be stored and retrieved using bichromatic electromagnetic induced transparency. 2\,MHz sidebands of squeezed vacuum pulses with temporal widths of 470\,ns and a squeezing level of $-1.78 \pm 0.02$\,dB were stored for 3\,$\mu$s in the laser-cooled ${}^{87} \mathrm{Rb}$ atoms. $-0.44 \pm 0.02$\,dB of squeezing was retrieved, which is the highest squeezing ever reported for a retrieved pulse.  
\end{abstract}

\pacs{42.50.Dv, 42.50.Gy}
\keywords{Suggested keywords}
\maketitle
Quantum information processing (QIP) for continuous variables of light has been developed that employs a squeezed vacuum whose frequency sidebands are entangled with each other \cite{Teleportation,EntanglementSwapping,DenceCoding}. The sidebands which are a few megahertz from the carrier frequency have been dealt with a variety of protocols, such as the first demonstration of quantum teleportation \cite{TeleportationExperiment}, to prevent the signal from being polluted by environmental noise. Meanwhile, non-Gaussian operation has been demonstrated by subtracting a photon from the squeezed vacuum in a certain frequency bandwidth around degenerate components \cite{OddPhotonNumberStates}. Accessibility to both the degenerate and sideband frequency components has been critical to the development of QIP for continuous variables of light. 

Quantum memory is crucial for the further development of QIP with light. However, quantum memories of a squeezed vacuum that have been demonstrated by utilizing the dark states of electromagnetically induced transparency (EIT) so far are only for degenerate components \cite{Storage,Memory} since conventional EIT creates only a single transparency window. Increasing the intensity of the control light, can broaden the transparency window so that it includes the high frequency sidebands, but this reduces the storage efficiency. In this Letter, we demonstrate a novel quantum memory that enables arbitrary frequency sidebands of a squeezed vacuum to be stored by using bichromatic control light. This method provides quantum memory that is completely compatible with the present QIP with continuous variables of light.

\begin{figure}[ht]
	\centering
	\includegraphics[width=\linewidth]{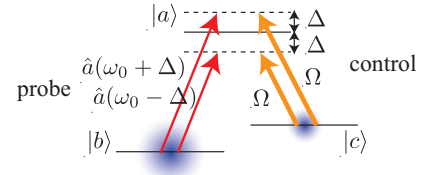}
	\vspace*{-5mm}
	\caption{(color online). Three-level atoms coupled to two-mode probe light and bichromatic control light.}
\end{figure}
Consider bichromatic EIT shown in Fig. 1. Two-mode probe light denoted by the annihilation operator $\hat{a}(\omega_{0} + \Delta)$ and $\hat{a}(\omega_{0} - \Delta)$ that are $\pm \Delta$ from the resonant frequency couple to the $\ket{b} \rightarrow \ket{a}$ transition of each of N atoms. Bichromatic control light detuned by $\pm \Delta$, both having Rabi frequencies of $\Omega$, couple to the $\ket{c} \rightarrow \ket{a}$ transition. The interaction Hamiltonian of the system is described in an interaction picture by
\begin{align}
	\hat{V} =& \hbar g \sum_{j = 1}^{N} \hat{a}(\omega_{0} + \Delta) e^{-i \Delta t} \hat{\sigma}_{ab}^{j} + \hbar g \sum_{j = 1}^{N} \hat{a}(\omega_{0} - \Delta) e^{i \Delta t} \hat{\sigma}_{ab}^{j} \nonumber \\
	&- \hbar \Omega e^{-i \Delta t} \sum_{j = 1}^{N} \hat{\sigma}_{ac}^{j} - \hbar \Omega e^{i \Delta t} \sum_{j = 1}^{N} \hat{\sigma}_{ac}^{j} + \mathrm{H. \, c.}, \label{eq:interaction hamiltonian 2mode}
\end{align}
where $\hat{\sigma}^{j}_{\mu \nu}$ is the flip operator of the $j$th atom and $g$ is the coupling constant between the atoms and the probe light.
To clarify the characteristics of this bichromatic EIT system, we introduce the following bosonic operators 
\begin{align}
	\hat{a}_{\pm} = \frac{1}{\sqrt{2}} [\hat{a}(\omega_{0} + \Delta) \pm \hat{a}(\omega_{0} - \Delta)]. 
\end{align}
The physical meaning of the modes $\hat{a}_{\pm}$ can be understood from their continuous-mode descriptions, i.e., 

\begin{align}
	\hat{a}_{+} = \frac{1}{\sqrt{\pi}} \int_{-\infty}^{\infty} \! d t \, \hat{a}(t) e^{i \omega_{0} t} \cos(\Delta t),\nonumber\\	
	\hat{a}_{-} = \frac{1}{\sqrt{\pi}} \int_{-\infty}^{\infty} \! d t \, \hat{a}(t) e^{i \omega_{0} t} \sin(\Delta t), \label{eq:a+}
\end{align}
where $\hat{a}(t)$ is the Fourier transformed operator of $\hat{a}(\omega)$ defined by $\hat{a}(t) = (2 \pi)^{-1/2} \int^{\infty}_{-\infty} d \omega  \hat{a}(\omega) \exp(-i \omega t)$. It is noted that the temporal function $\cos(\Delta t)$ describing the mode $\hat{a}_{+}$ is identical to the beating of the amplitude of the control light due to their two frequency components interfering.

Using these bosonic operatos, Eq. (\ref{eq:interaction hamiltonian 2mode}) is rewritten as
\begin{align}
	\hat{V} =& \sqrt{2} \hbar g \cos(\Delta t) \sum_{i = 1}^{N} \hat{a}_{+} \sigma_{ab}^{i} - 2 \hbar \Omega \cos(\Delta t) \sum_{i = 1}^{N} \sigma_{ac}^{i} \nonumber \\
	&+ \sqrt{2} \hbar g \sin(\Delta t) \sum_{i = 1}^{N} \hat{a}_{-} \sigma_{ab}^{i} + \mathrm{H. \, c.}. \label{eq:interaction hamiltonian}
\end{align}
Since, the first and the second terms which indicate the interaction between the atoms and the mode $\hat{a}_{+}$ and that between the atoms and the control light have the same time dependence of $\cos(\Delta t)$, there exists dark states for the mode $\hat{a}_{+}$. The presence of dark states permits EIT as well as monochromatic EIT \cite{DarkStatePolaritons}. In contrast, the interaction between the atoms and the mode $\hat{a}_{-}$ has a time dependence of $\sin(\Delta t)$ and there are no dark states for the mode $\hat{a}_{-}$. The probe can thus perfectly transmit atoms only when the probe is the excitation of the mode $\hat{a}_{+}$.

In the following, we describe the relation between the squeezed vacuum and the modes $\hat{a}_{\pm}$. We treat the two-mode squeezed vacuum of the sideband frequency of $\Delta$ defined as
\begin{align}
	\ket{\zeta} = \exp[ & \zeta^{\ast} \hat{a}(\omega_{0} + \Delta) \hat{a}(\omega_{0} - \Delta) \nonumber \\
	&-\zeta \hat{a}^{\dagger}(\omega_{0} + \Delta) \hat{a}^{\dagger}(\omega_{0} - \Delta)] \ket{0}, \label{eq:2mode squeezed vacuum}
\end{align}
where $\zeta$ is the complex squeezing parameter. This can be written in terms of $\hat{a}_{\pm}$ as 
\begin{align}
	\ket{\zeta} = \exp \left( \frac{\zeta^{\ast} \hat{a}^{2}_{+}}{2} - \frac{\zeta (\hat{a}_{+}^{\dagger})^{2}}{2} \right) \exp \left( - \frac{\zeta^{\ast} \hat{a}^{2}_{-}}{2} + \frac{\zeta (\hat{a}_{-}^{\dagger})^{2}}{2} \right) \ket{0}.
\end{align}
A two-mode squeezed vacuum is thus represented by the product of the single-mode squeezed vacua at the modes $\hat{a}_{\pm}$. 
That is, when bichromatic control light are employed, only the squeezed vacuum for the mode $\hat{a}_{+}$ can be stored in atoms.

In experiments involving the sidebands of the squeezed vacuum, homodyne detection with a local oscillator (LO) light, the frequency of which is identical to the frequency of the squeezed vacuum, is employed and the power of the frequency component corresponding to the sideband frequency is measured. The obtained power is represented by $\braket{\hat{X}_{\Delta}^{\dagger}(\theta) \hat{X}_{\Delta}(\theta)}$, where $\hat{X}_{\Delta}(\theta) = [\hat{a}^{\dagger}(\omega_{0} - \Delta) e^{i \theta} + \hat{a}(\omega_{0} + \Delta) e^{-i \theta}]/2$ is the two-mode quadrature and $\theta$ is the relative phase between the squeezed vacuum and the LO. This power can be rewritten with the quadratures of the modes $\hat{a}_{\pm}$, which are defined as $\hat{X}_{\pm} = (\hat{a}_{\pm}^{\dagger} e^{i \theta} + \hat{a}_{\pm} e^{-i \theta})/2$, i.e.,
\begin{align}
	\braket{\hat{X}_{\Delta}^{\dagger}(\theta) \hat{X}_{\Delta}(\theta)} = \frac{1}{2} \braket{\hat{X}_{+}^{2}(\theta)} + \frac{1}{2}\braket{\hat{X}_{-}^{2}(\theta + \pi/2)}. \label{eq:homodyne}
\end{align}
Since the both $\hat{a}_{+}$ and $\hat{a}_{-}$ modes contribute to Eq. (\ref{eq:homodyne}), the optical loss for the mode $\hat{a}_{-}$ degrades the observable squeezing. However, since the  temporal function describing the mode $\hat{a}_{+}$ is already known as is shown in Eq. (\ref{eq:a+}), employing the time domain homodyne method \cite{SinCosQND, TemporalFunction}, we can evaluate only the quadrature of the mode $\hat{a}_{+}$.

\begin{figure}[ht]
	\begin{center}
		\includegraphics[width=\linewidth]{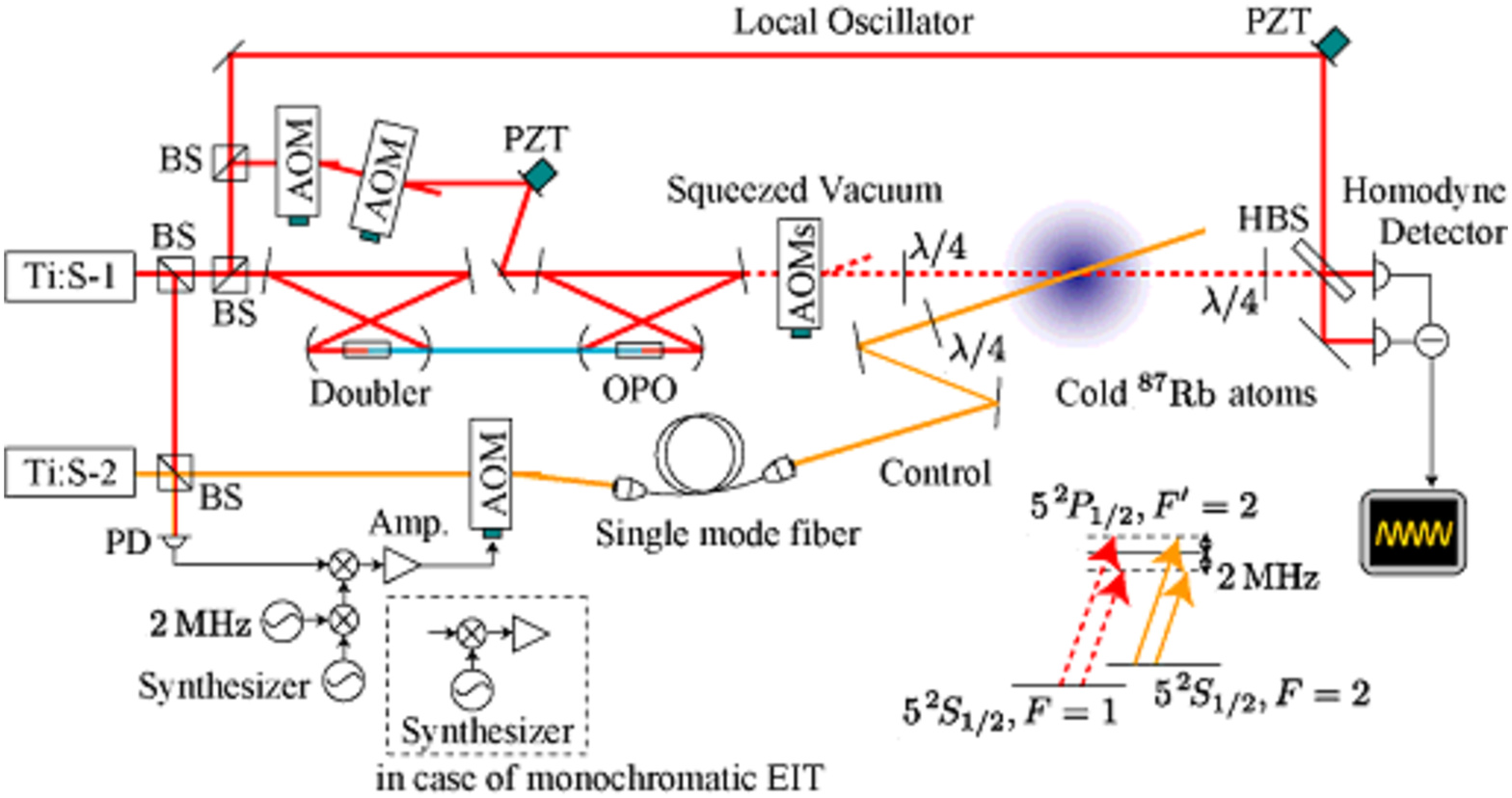}
		\caption{(color online). Schematic diagram of the experimental setup. BS: beam splitter, HBS: half beam splitter, AOM: acousto-optic modulator, PD: photodetector, PZT: piezo electric transducer, Amp: RF amplifier.}
	\end{center}
\end{figure}
Our experimental setup is shown schematically in Fig. 2. A laser-cooled atomic ensemble of ${}^{87} \mathrm{Rb}$ was utilized as the EIT medium. Ti:sapphire laser 1 was employed for generating and detecting the squeezed vacuum whose carrier frequency was resonant on the $5 \,{}^{2} S_{1/2}, F = 1 \rightarrow 5 \,{}^{2} P_{1/2}, F^{\prime} = 2$ transition. Ti:sapphire laser 2 was diffracted by an acousto-optic modulator (AOM) and was used as the control light. The relative frequency between the probe and the control light was stabilized using the feed-forward method \cite{FeedForward}. By adjusting the frequency of the synthesizer, the detuning of the control light could be precisely controlled around the $5 \,{}^{2} S_{1/2}, F = 2 \rightarrow 5 \,{}^{2} P_{1/2}, F^{\prime} = 2$ transition. When bichromatic control light were used, the synthesizer output was mixed with a sine wave at the frequency $\Delta$ and the bichromatic control light were detuned by $\pm \Delta$. Although the diffraction angles of the AOM for these two frequency components differed slightly, they were coupled to the single mode fiber and thus had identical spatial modes when they were injected into the cold atoms. The probe and control light were circularly polarized in the same direction and were incident on the cold atoms with a crossing angle of $2.5^{\circ}$. The radii of the probe and control light were 170\,$\mu$m and 790\,$\mu$m, respectively.  
One cycle of our experiment was 10\,ms. Each cycle consisted of a 9\,ms preparation period to prepare the cold atoms in the $5 \,{}^{2} S_{1/2}, F = 1$ state and a 1\,ms measurement period (for details, refer to \cite{TimeDomainHomodyne}). The optical depth of the atoms was approximately eight.

We first demonstrate experimentally that EIT cannot be achieved for high frequency sidebands when a monochromatic control light is utilized. During the measurement period, the squeezed vacuum was incident on the cold atoms with the control light in a coherent state and having a power of 200\,$\mu$W. The squeezed vacuum passed through the cold atoms and was measured by the time-domain homodyne method. The measured homodyne signals were imported into the high-speed digital oscillator, which had a signal sampling rate of 5$\times 10^7$\,samples/sec. We evaluated the spectrum of the quadrature noise by taking the Fourier transform of the obtained data and squaring it. The relative phase between the squeezed vacuum and the LO was locked to $\theta = \pi/2$ or $0$ in the preparation period by using a weak coherent beam, which had an identical spatial mode as that of the squeezed vacuum. During the measurement period, the feedback voltages to PZTs were kept constant and this beam was turned off using the AOM. 

\begin{figure}[ht]
	\begin{center}
		\includegraphics[width=\linewidth]{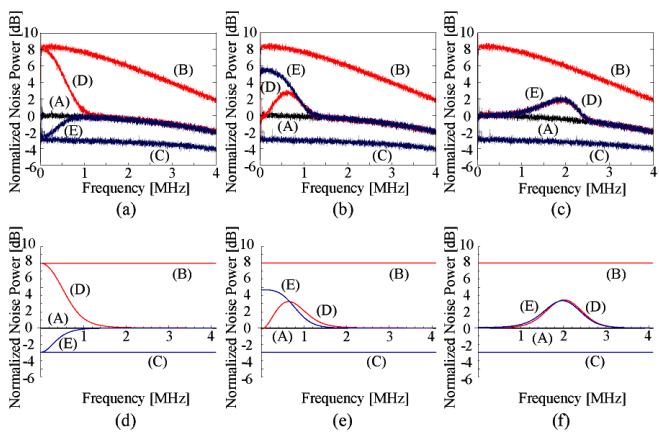}
		\caption{(color online). Quadrature noise spectra of the squeezed vacuum incident together with the monochromatic control light, where the frequencies of the control light were (a) resonant on the $5 \,{}^{2} S_{1/2}, F = 2 \rightarrow 5 \,{}^{2} P_{1/2}, F^{\prime} = 2$ transition, (b) detuned by 500\,kHz and (c) detuned by 2\,MHz. Figures 3(d), (e) and (f) show the numerical simulations of the EIT for the squeezed vacuum shown in Figs. 3(a), (b) and (c), respectively.}
	\end{center}
\end{figure}
Fig. 3(a) shows the observed spectra of the quadrature noise when the frequency of the control light was resonant on the $5 \,{}^{2} S_{1/2}, F = 2 \rightarrow 5 \,{}^{2} P_{1/2}, F^{\prime} = 2$ transition. Trace (A) indicates the shot noise and traces (B) and (C) indicate the quadrature noise of the squeezed vacuum in the absence of cold atoms. Traces (D) and (E) indicate the quadrature noise of the squeezed vacuum that passed through the cold atoms due to EIT. Here, the relative phases were $\pi/2$ in traces (B) and (D), and 0 in traces (C) and (E). Each data point is an average of 1000 measurements. EIT for the squeezed vacuum was observed in the low frequency region, and the cold atoms were made almost transparent for the degenerate components of the squeezed vacuum. When the control light is resonant on the transition, each sideband constructing the two-mode squeezed vacuum undergo phase shifts of the opposite sign, and thus the effect of the phase shifts is canceled \cite{TimeDomainHomodyne}. Fig. 3(d) shows the corresponding numerical simulation, where the atomic absorption loss and the phase shift caused by dispersion under the EIT condition were taken into account.

Figs. 3(b) and 3(e) show the experimental results and numerical simulations when the control light was detuned by 500\,kHz. Since there is no longer perfect cancellation of the dispersion effect, the noise spectra is completely different. With further detuning of the control light (Fig. 3(c), where the control light was detuned by 2\,MHz), only one of the sidebands constructing the two-mode squeezed vacuum passed through the atoms, while the other was absorbed. The observed quadrature noise thus exceeded the vacuum noise and was phase insensitive. The corresponding numerical simulation is shown in Fig. 3(f). These results demonstrate that EIT cannot be achieved for the high frequency sidebands of a squeezed vacuum by simply detuning the control light. 

Next, we demonstrate EIT for a squeezed vacuum using bichromatic control light. During the measurement period, the squeezed vacuum and bichromatic control light, which were detuned by $\pm$2\,MHz, were incident on the cold atoms. The frequency components of the control light have the same power, giving a total power of 200\,$\mu$W. Fig. 4(a) shows the quadrature noise spectra evaluated by taking the Fourier transform of the homodyne signals and squaring them, as was done in Fig. 3. In contrast to Fig. 3, both antisqueezing and squeezing were observed around 2\,MHz. However, as has already been discussed, transmission was limited to 50\%{} due to absorption for the mode $\hat{a}_{-}$.
\begin{figure}[ht]
	\begin{center}
		\includegraphics[width=\linewidth]{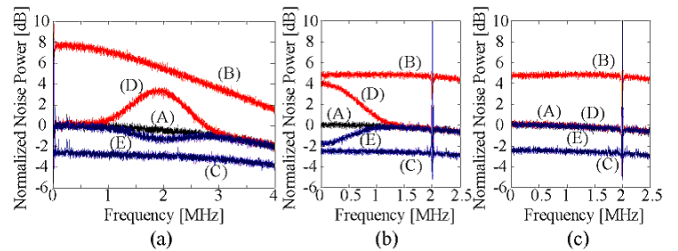}
		\caption{(color online). Quadrature noise spectra of the squeezed vacuum incident with bichromatic control light detuned by $\pm$2\,MHz obtained by (a) taking the Fourier transform of direct homodyne signals and squaring them, and obtained by (b) multiplying the 2\,MHz sine waves having phases that are identical to that of the beating amplitude of the control light and (c) that differed by $\pi/2$ from the beating amplitude of the control light before taking the Fourier transform.}
	\end{center}
\end{figure}

In order to extract the quadrature noise of the mode $\hat{a}_{+}$, we need information about the phase of the beating amplitude of the control light. In our setup, this phase was determined by that of the 2\,MHz sine wave mixed with the synthesizer output. We measured this sine wave together with the homodyne signal, which enabled us to determine the temporal function describing the mode $\hat{a}_{+}$. The obtained data of the homodyne signal was multiplied by this 2\,MHz sine wave of the temporal function, and then the power spectrum was evaluated (Fig. 4(b)). The transparency window, which appeared around 2\,MHz in Fig. 4(a), was down-converted to the low frequency region and the transparency was greatly enhanced (approximately 75\%{}) compared that of Fig. 4(a). Moreover, the environmental noise in the low frequency region in Fig. 4(a) were up-converted to about 2\,MHz, and they did not appear in the transparency window, in contrast with Fig. 3(a). 
The quadrature noise of the mode $\hat{a}_{-}$ was observed (Fig. 4(c)) by utilizing the 2\,MHz sine wave whose phase differed by $\pi/2$ from that of the beating amplitude of the control light. The squeezed vacuum was completely absorbed, as theoretically predicted.

\begin{figure}[ht]
	\begin{center}
		\includegraphics[width=0.9 \linewidth]{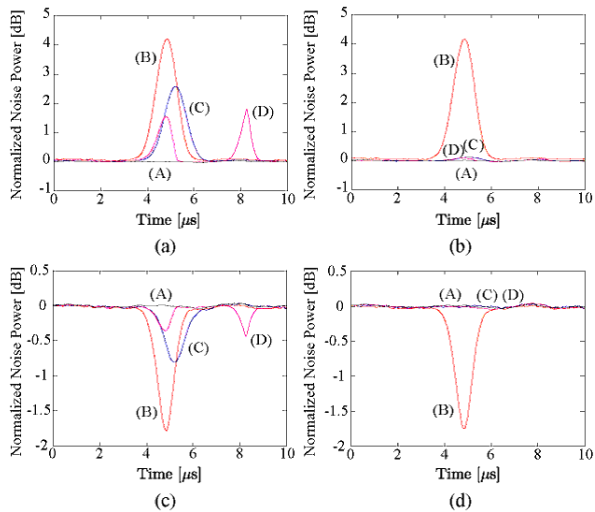}
		\caption{(color online). Temporal variation of the quadrature noises of the mode $\hat{a}_{+}$ (left column) and the mode $\hat{a}_{-}$ (right column), where relative phases between the squeezed vacuum and the LO were set to $\pi/2$ ((a) and (b)) and $0$ ((c) and (d)).}
	\end{center}
\end{figure}

Finally, we demonstrated storage and retrieval of the squeezed vacuum with bichromatic EIT. Squeezed vacuum pulses, which were Gaussian with temporal widths of 470\,ns, were created from the continuous-wave squeezed vacuum with three AOMs. The residual photon flux at the tail of the pulses was approximately 1\%{}. The pulsed squeezed vacuum was incident on the cold atoms together with bichromatic control light and it was stored in the atoms by turning off the control light using the AOM. It was retrieved after 3\,$\mu$s by turning the control light on. The homodyne signal data were acquired at a signal sampling rate of 1$\times 10^8$\,samples/sec. In order to observe the squeezed vacuum pulses at the mode $\hat{a}_{+}$, the obtained data were multiplied by a 2\,MHz sine wave and then multiplied by a temporal function corresponding to the retrieved pulse. Finally, they were integrated and squared to evaluate the temporal variation of the quadrature noise of the mode $\hat{a}_{+}$. The temporal function employed here was a half-Gaussian expected from experimental results of storage and retrieval of the pulses in a coherent state. The quadrature noise of the mode $\hat{a}_{+}$ is shown in Figs. 5(a) ($\theta=\pi/2$) and 5(c) ($\theta=0$), and that of the mode $\hat{a}_{-}$ is shown in Figs. 5(b) ($\theta=\pi/2$) and 5(d) ($\theta=0$). Traces (A) indicate the shot noise and the traces (B) and (C) indicate the quadrature noise of the squeezed vacuum pulses with no cold atoms and the ultraslow propagation of squeezed vacuum pulses with EIT, respectively. Traces (D) indicate the quadrature noise when the squeezed vacuum pulses were stored and retrieved. Traces (A), (B) and (D) were averaged 900 000 times, and traces (C) were averaged 450 000. The observed antisqueezing and squeezing levels for the retrieved squeezed vacuum pulses were 1.80$\pm$0.02\,dB and -0.44$\pm$0.02\,dB, respectively. In the present experiment, the quadrature noise was evaluated in a frequency region free from environmental noise, but it slightly mixed in the signals. The shot noise thus fluctuated slightly and the margin of error was estimated by the standard deviation of the temporal variation of the shot noise.

In summary, we demonstrated a novel type of quantum memory that can be applied to arbitrary sidebands of a squeezed vacuum. This method is robust against environmental noise and is also compatible with current QIP for continuous variables of light.

We would like to express our sincere gratitude to K. Akiba and Y. Yokoi for valuable comments and stimulating discussions. This work was supported by the Global Center of Excellence Program by MEXT, Japan through the Nanoscience and Quantum Physics Project of the Tokyo Institute of Technology, and by MEXT through a Grant-in-Aid for Scientific Research (B).

\end{document}